\documentclass[aps,pra,showpacs,twocolumn,
amsmath,superscriptaddress]{revtex4-1}
\usepackage{pifont}
\usepackage{graphicx,epsfig,subfigure,dsfont,amssymb,amsmath,amsthm,amsfonts,amsbsy,mathrsfs,amscd}
\usepackage{epstopdf}
\usepackage{array}
\usepackage{booktabs}
\usepackage{multirow}

\usepackage{mathrsfs}

\DeclareMathOperator{\Span}{span}
\DeclareMathOperator{\tr}{tr}
\DeclareMathOperator{\rank}{rank}
\DeclareMathAlphabet{\mathpzc}{OT1}{pzc}{m}{it}

\newtheorem{theorem}{Theorem}

\usepackage[unicode=true,
 bookmarks=true,bookmarksnumbered=false,bookmarksopen=false,
 breaklinks=false,pdfborder={0 0 1},backref=false,colorlinks=true]
 {hyperref}
\hypersetup{
 linkcolor=magenta, urlcolor=blue, citecolor=blue, pdfstartview={FitH}, hyperfootnotes=false, unicode=true}

\usepackage{color}
\newcommand{\tabincell}[2]{\begin{tabular}{@{}#1@{}}#2\end{tabular}}

\def\la{\langle}
\def\ra{\rangle}

\begin{document}
\title{Necessity for quantum coherence of nondegeneracy in energy flow}

\smallskip
\author{Teng Ma}
\affiliation{Shenzhen Institute for Quantum Science and Engineering and Department of Physics, Southern University of Science and Technology, Shenzhen 518055, China}
\affiliation{Key Laboratory of Quantum Information,
University of Science and Technology of China, CAS, Hefei 230026, China}
\affiliation{Shenzhen Key Laboratory of Quantum Science and Engineering, Southern University of Science and Technology, Shenzhen 518055, China}

\author{Ming-Jing Zhao}
\affiliation{School of Science,
Beijing Information Science and Technology University, Beijing 100192, China}

\author{Shao-Ming Fei}
\email{feishm@cnu.edu.cn}
\affiliation{School of Mathematical Sciences, Capital Normal
University, Beijing 100048, China}
\affiliation{Max-Planck-Institute for Mathematics in the Sciences,  Leipzig 04103, Germany}

\author{Man-Hong Yung}
\email{yung@sustc.edu.cn}
\affiliation{Shenzhen Institute for Quantum Science and Engineering and Department of Physics, Southern University of Science and Technology, Shenzhen 518055, China}
\affiliation{Shenzhen Key Laboratory of Quantum Science and Engineering, Southern University of Science and Technology, Shenzhen 518055, China}
\affiliation{Central Research Institute, Huawei Technologies, Shenzhen, 518129, China}


\begin{abstract}
In this work, we show that the quantum coherence among non-degenerate energy subspaces (CANES for short) is essential for the energy flow in any quantum system. 
 CANES satisfies almost all of the requirements as a coherence measure, except that the coherence within degenerate subspaces is explicitly eliminated. We show that the energy of a system becomes frozen if and only if the corresponding CANES vanishes, which is true regardless the form of interaction with the environment. However, CANES can remain zero even if the entanglement changes over time. Furthermore, we show how the power of energy flow is bounded by the value of CANES. An explicit relation connecting the variation of energy and CANES is also presented. These results allow us to bound the generation of system-environment correlation through the local measurement of system's energy flow.
\end{abstract}

\maketitle


\section{Introduction}

The transfer of energy from one system to another is one of the most fundamental processes in nature. From the modern point of view, the flow of energy is always associated with the flow of information, or variation of correlations,  between the physical systems, which has  become an active research area in the context of quantum thermodynamics~ \cite{kawsaki,mcgaughey,valla,Haedler,lloyd,Micadei,Henao,Latune,Elouard,alvaro,rosario}. In particular, recent works on quantum battery~\cite{Campaioli,Ferraro,Alicki,Campaioli2} and quantum heat engine~\cite{Klatzow,Uzdin} suggest that quantum coherence can provide advantages over many tasks.

On the other hand, a quantum mechanical understanding of energy flow is crucial for quantum biology. In particular, for photosynthetic light-harvesting complexes~\cite{Lambert,cheng}, it has been speculated that quantum coherence may play an important role in boosting the efficiency of energy transfer in this biological system~\cite{Engel,Lee}, even though the photosynthetic complexes are working in a ``hot-and-wet" environment~\cite{Plenion,mohseni,Rebentrost}.


As one of the key features in quantum theory, quantum coherence is indispensable for many physical phenomena and applications, such as interference of light, laser~\cite{mandel}, and superconductivity~\cite{london}. Similar to quantum entanglement~\cite{horodecki} and other quantum correlations such as discord~\cite{olivier,discord,Henderson}, quantum coherence can be regarded as a kind of physical resource~\cite{Streletsov,Tbaum,radhakri,chuantan,guoyu,jiajun ma,Marvian,Mondal,xksong} for quantum computation and quantum information processing~\cite{nielsen},  and also in thermodynamic processes~\cite{oppenheim,horodeckioppenheim}.
Recent work of quantifying coherence \cite{Tbaum} in the context of quantum information science gives an informational perspective for coherence. Moreover, the coherence is found to be connected to the quantum correlations \cite{radhakri,chuantan,guoyu,jiajun ma,slluo,shkim}.

\begin{figure}[b!]
\centering
\includegraphics[width=\columnwidth]{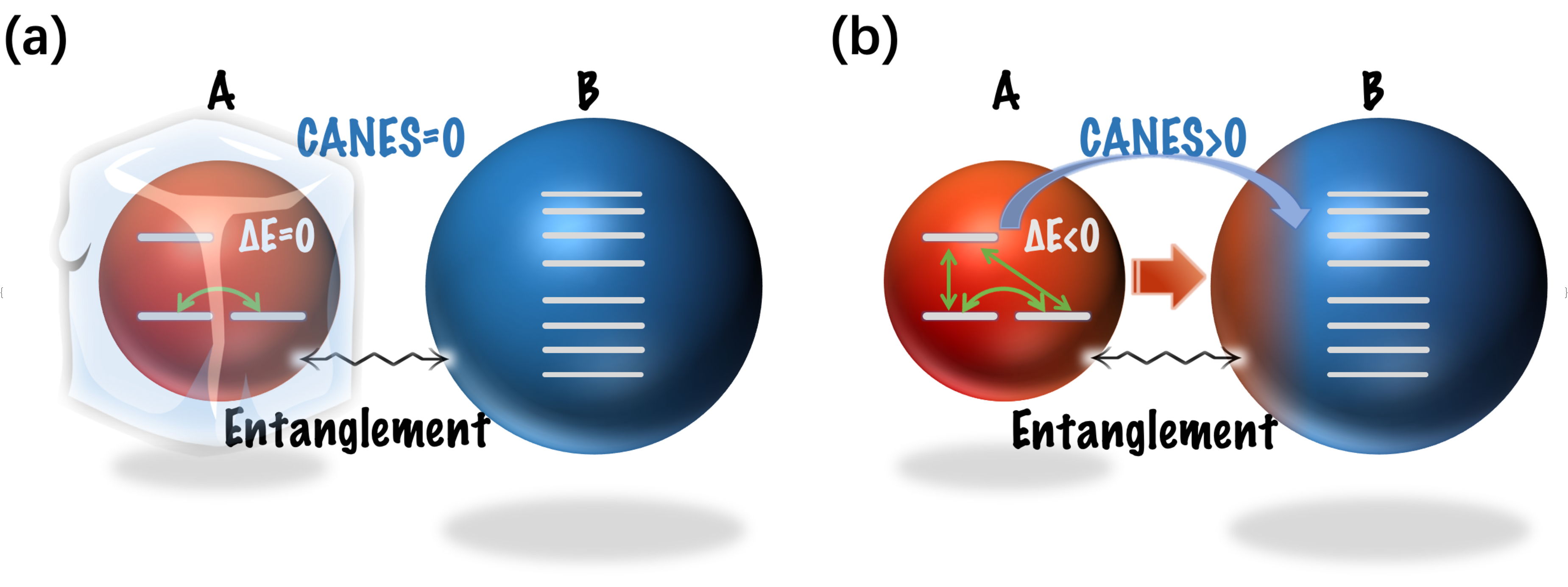}
\hspace{0.01in}
\caption{The role played by CANES in energy flow. (a) The local average energy of system A  keeps a constant for any interaction iff the corresponding CANES gets zero.  (b) When there is some energy flow from system A, the CANES limits the power of energy flow. The system A and its environment B may be entangled  due to the degenerated energy levels of system A for both cases.}
\label{energy_carton}
\end{figure}

In this work, we uncover the role of quantum coherence in energy flow between two quantum systems, from the quantum information perspective. The key physical quantity is called {\it coherence among non-degenerate energy subspaces} (or ``CANES" for short), which measures the quantum coherence among different eigen-subspaces of a given Hamiltonian.
Specifically, we show that CANES imposes an upper bound for the power of the energy flow (i.e., the change rate of energy). Furthermore, we prove that the vanishing of CANES is both sufficient and necessary for freezing the energy flow of open quantum systems. In addition, we also establish a CANES production-energy flow relation for non-equilibrium systems, which limits the energy variation in terms of CANES. An example is given for the purpose of illustration.

\section{CANES}
Let us consider a physical system with Hamiltonian~$H=\sum\nolimits_i {{h_i}} {P_i}$, where $h_i$ is the $i$-th eigenvalue which generally associates to a degenerate subspace spanned by $\Span \left\{ {\left| {{e_{i1}}} \right\rangle ,\left| {{e_{i2}}} \right\rangle ,\left| {{e_{i3}}} \right\rangle ...} \right\}$, ${P_i} \equiv \sum\nolimits_k {\left| e_{ik} \right\rangle \left\langle e_{ik} \right|}$ whose rank is equal to the degeneracy of the eigenvalue $h_i$. For any state~$\rho$, the CANES is defined as follows:
\begin{equation}\label{ec}
\mathcal{C}_{e}(\rho) \equiv S(\Lambda \rho)-S(\rho) \, ,
\end{equation}
where  $S(\rho) \equiv -\tr(\rho \ln \rho)$ is the von Neumann entropy, $\Lambda =\sum_i P_i(\cdot)P_i$ is a superoperator.
The CANES reduces to the original relative-entropy coherence \cite{Tbaum} only if the Hamiltonian $H$ is non-degenerate. 
It should be pointed out that the CANES is actually within the definition of the relative entropy of superposition appeared in \cite{JAberg}: the CANES is respect to specific spaces, i.e., the energy subspaces, while the relative entropy of superposition can be with respect to any spaces. 

Furthermore, the CANES vanishes if and only if the coherence among different eigen-subspaces vanishes, i.e.,
\begin{equation}
\mathcal{C}_{e}(\rho)=0\ \Leftrightarrow \ \rho =\sum_i P_i \rho P_i \, .
\end{equation}
In addition, since $\rho  = \sum\nolimits_i {{P_i}} \rho {P_i} \Leftrightarrow \left[ {\rho ,H} \right]=0$, see Appendix \ref{Proper_CANES}, we can also conclude that the state $\rho$ becomes stationary, i.e., $\dot \rho  = i\left[ {\rho ,H} \right] = 0$ (setting $\hbar=1$), which also implies that
\begin{equation}\label{CiffrhoH}
\mathcal{C}_{e}(\rho)=0 \ \Leftrightarrow \  [\rho, H]=0 \, .
\end{equation}

The concept of CANES can be extended to bipartite systems $A$ and $B$. Denote the local Hamiltonians by $H_X=\sum_i h^X_i P^X_i$ ($X=A, B$),
where $P^X_i \equiv \sum\nolimits_k {\left| e^X_{ik} \right\rangle \left\langle e^X_{ik} \right|}$ is
the projector onto the eigen-subspace corresponding to the eigenvalue $h^X_i$.
 For a given bipartite state $\rho_{AB}$, we define the left and right CANES as,
 $\mathcal{C}^{\rightarrow}_{e}(\rho_{AB})  \equiv S((\Lambda_A\otimes\mathbb{I})\rho_{AB})-S(\rho_{AB})$,
 $\mathcal{C}^{\leftarrow}_{e}(\rho_{AB}) \equiv S(\mathbb{I} \otimes \Lambda_B) \rho_{AB})-S(\rho_{AB})$,
where $\Lambda_X \equiv \sum_iP^X_i(\cdot)P^X_i$.
Furthermore, we will also need to define the bilateral CANES, $\mathcal{C}^{\leftrightarrow}_{e}(\rho_{AB}) \equiv S((\Lambda_A\otimes \Lambda_B)\rho_{AB})-S(\rho_{AB})$.


\section{Freezing theorem for local energy}

To show that CANES is closely related to the energy flow between two quantum sub-systems A and B, subjecting to the global Hamiltonian, $H=H_A+H_B+H_I$ with time-independent $H_X$, let us consider the Liouville-von Neumann equation of motion, $i\frac{d}{d t}\rho_{AB}=[H,\rho_{AB}]$. The local energies ${E_X} \equiv {\rm tr}\left( {{\rho_{AB}}{H_X}} \right)$ satisfy the following equation of motion,
\begin{equation}\label{dEdttrtr}
i\frac{d}{{dt}}{E_X} = {\text{tr}}\left( {[H,\rho _{AB}] H_X} \right) = {\text{tr}}\left( {{H_I}\left[ {{\rho _{AB}},{H_X}} \right]} \right) \, .
\end{equation}
A generalization of Eq.~(\ref{CiffrhoH}), e.g.,
\begin{equation}\label{gen-Crhiff}
\mathcal{C}^{\rightarrow}_e \left( {{\rho _{AB}}} \right) = 0 ~ \Leftrightarrow ~ \left[ {{\rho _{AB}},{H_A}} \right] = 0\,,
\end{equation}
implies that the rate of local energy $\frac{d}{{dt}}{E_X} = 0$ for all $H_I$, if and only if $\mathcal{C}^{\rightarrow}_e \left( \rho _{AB} \right) = 0$ (for $X=A$) or $\mathcal{C}^{\leftarrow}_e \left( \rho _{AB} \right) = 0$ (for $X=B$). These results are summarized by the following theorem:

\begin{theorem}[{\bf Freezing of local energy}]\label{thmfrezenlo}
For a period of time $t\in [0,\tau]$, the local energy is a constant for any interaction $H_I$ if and only if the corresponding CANES remains zero, i.e.,
\begin{align}\label{frozen energy}
&\mathcal{C}^{\rightarrow}_{e}(\rho_{AB})=0
~ \Leftrightarrow ~ E_A={\rm constant}~\forall\, H_I \, , \\
&\mathcal{C}^{\leftarrow}_{e}(\rho_{AB})=0~ \Leftrightarrow ~ E_B={\rm constant} ~\forall\, H_I \, , \\
&\mathcal{C}^{\leftrightarrow}_{e}(\rho_{AB})=0  \ \Leftrightarrow \ E_A \; \& \; E_B={\rm constant} ~ \forall\, H_I \, .
\end{align}
\end{theorem}

The relations in Theorem 1 indicate the sufficiency and necessity of the CANES in energy flow. 
In particular, if there is energy flow between the system and its environment, the associated CANES must be non-zero.



\begin{table}[b!]
\caption{Implications to local CANES and quantum correlations, when local energies are frozen and the corresponding local Hamiltonians are non-degenerate (ND).}
\begin{tabular}{l|c|c|c}
\toprule[1pt]
\multicolumn{1}{c|}{} & \multicolumn{1}{c|}{$\mathcal{C}^{\rightarrow}_e(\rho_{AB})=0$} & \multicolumn{1}{c|}{$\mathcal{C}^{\leftarrow}_e(\rho_{AB})=0$} & \multicolumn{1}{c}{$\mathcal{C}^{\leftrightarrow}_e(\rho_{AB})=0$} \\
\hline
\multicolumn{1}{c|}{Local CANES}    &     $\mathcal{C}_e(\rho_A)=0$ & $\mathcal{C}_e(\rho_B)=0$     &      \tabincell{c}{$\mathcal{C}_e(\rho_A)=0$\\$\mathcal{C}_e(\rho_B)=0$}     \\
\hline
 $H_A$ ND                &   I-Q state\footnote{Incoherent-quantum state, see main text.}                     &                 &                   \\
\hline
 $H_B$     ND            &                       &      Q-I state\footnote{Similar to I-Q state, with A and B swapped.}                 &                     \\
\hline
 $H_A \& H_B$ ND           &       I-Q state                &      Q-I state                 &         I-I state\footnote{Incoherent-Incoherent state, see main text.}               \\
 \bottomrule[1pt]
\end{tabular}
\label{tab:2}
\end{table}

In fact, there are two more consequences when the bipartite CANES vanishes.
(i) The corresponding local CANES is zero, i.e.,
\begin{equation}\label{ulczimpcer}
\mathcal{C}^{\rightarrow}_e\left( {{\rho _{AB}}} \right) = 0~\Rightarrow ~ {\mathcal{C}_e}\left( {{\rho _A}} \right) = 0 \, ,
\end{equation}
where $\rho_A \equiv \tr_B (\rho_{AB})$. It is because, for any local operator $O_A \otimes I$, the following relation holds, $\left[ {{\rho _{AB}},{O_A} \otimes I} \right] = 0 \Rightarrow \left[ {{\rho _A},{O_A}} \right] = 0$ (the opposite may not be true). Therefore, we have $\left[ {{\rho _A},{H_A}} \right] = 0$ from Eq.~(\ref{gen-Crhiff}), which implies the advertised result in Eq.~(\ref{ulczimpcer}) from Eq.~(\ref{CiffrhoH}). Furthermore, since  $\mathcal{C}_e^ \leftrightarrow  \left( {{\rho _{AB}}} \right) = 0 \Leftrightarrow \mathcal{C}^{\leftarrow}_{e}(\rho_{AB})=0~\&~\mathcal{C}^{\rightarrow}_{e}(\rho_{AB})=0$ (see Appendix \ref{Proper_CANES}), we can also conclude that
\begin{equation}
\mathcal{C}_e^\leftrightarrow  \left( {{\rho _{AB}}} \right)=0 \ \Rightarrow \ {\mathcal{C}_e}\left( {{\rho _A}} \right) = 0\; \& \; {\mathcal{C}_e}( {{\rho _B}} )=0 \, .
\end{equation}
(ii) When $H_A$ is non-degenerate, $\mathcal{C}^{\rightarrow}_{e}(\rho_{AB})=0$ implies (see Appendix \ref{Proper_CANES}) that the state can be written as an incoherent-quantum state~\cite{strltsov,mateng}, i.e.,
\begin{equation}
\mathcal{C}^{\rightarrow}_e\left( {{\rho _{AB}}} \right) = 0 \ \Rightarrow  \ {\rho _{AB}} = \sum\limits_i {{p_i}} \left| {{e^A_i}} \right\rangle \left\langle {{e^A_i}} \right| \otimes \rho _i^B \, ,
\end{equation}
where $|e^A_i\ra$s are the
 eigenbasis of $H_A$.
Furthermore, when $H_A$ and $H_B$ are both non-degenerate, $\mathcal{C}^{\leftrightarrow}_{e}(\rho_{AB})=0$ implies the state can be written as an incoherent-incoherent state~\cite{strltsov,mateng}, i.e., $\rho_{AB} =\sum_{ij} p_{ij} |e^A_i\ra \la e^A_i| \otimes  |e^B_j\ra \la  e^B_j|$. In this case,
 since continuous unitary evolution does not change the eigenvalues
 $p_{ij}$ of $\rho_{AB}$, the state becomes invariant over time.
Hence freezing a system's energy imposes restrictions not  only on the system's  coherence,
 but also on the way that the system correlates with its environment (see  Table~\ref{tab:2}).

 It should be noted that, in order to keep CANES vanishing for $t \in [0, \tau]$, strong restrictions on $H_I$ in Theorem 1 need to be imposed, although there are no such restrictions on $H_I$ for the rate of energy flow $\frac{d}{{dt}}{E_X}=0$ in Eq. (\ref{dEdttrtr}). For example, keeping $\mathcal{C}_e^{\rightarrow}(\rho_{AB})=0$ implies that $\rho_{AB}$ keeps block diagonal under the product of local energy basis (the block scale is dependent on the degeneracy of $H_A$). And a sufficient condition  for that is that $H_I$ has the same shape (block diagonal form) as $\rho_{AB}$. Furthermore, when both $H_A$ and $H_B$ are non-degenerate, keeping $\mathcal{C}_e^{\leftrightarrow}(\rho_{AB})=0$ implies that $\rho_{AB}$ keeps diagonal under the product of local energy basis, for which a sufficient condition is that $H_I$ also has the diagonal form.




For bipartite systems, there is a related quantum correlation called discord \cite{olivier,discord,Henderson}, which can also be classified as left, right, and bilateral. For example, the left discord is defined as $\mathcal{D}^{\rightarrow}(\rho_{AB}) \equiv \min_{\Pi_A} \{I(\rho_{AB})-I(\Pi_A \rho_{AB})\}$, where $I(\rho_{AB}) \equiv S(\rho_{AB})+S(\rho_{B})-S(\rho_{AB})$ is the quantum mutual information associated with the von Neumann entropy. 
Here $\Pi_A$ is a set of rank-one POVM projectors on system A.
Recently, the role of discord in energy transport has been investigated~\cite{lloyd}. It was found that whenever the bilateral discord is zero, the joint system becomes effectively non-interacting. Here we found that whenever the local energies are frozen, the corresponding discord {\it may or may not} be zero. This observation is justified explicitly through an example shown in Appendix \ref{changeingcorrelation}, where we showed that both entanglement and discord can be non-zero, even if the local energies are frozen. Overall, instead of discord, our results show that CANES plays a necessary and sufficient role in freezing the local energies of quantum systems.


Moreover, we found that discord does play a role in freezing local energy when the corresponding local system is in a thermal state, i.e., for system A, $\rho_A^{\rm th}\propto e^{-\beta_A H_A}$, where $\beta_A$ is the inverse temperature. Note that system B is not necessarily in a thermal state. First of all, we found that zero left discord ${\mathcal{D}^ \to }({\rho _{AB}}|\rho_A^{\rm th})$ (conditioned on A being thermal) implies local energy freezing, i.e., ${\mathcal{D}^ \to }({\rho _{AB}}|\rho_A^{\rm th}) = 0 \Rightarrow {E_A}{\text{ }} = {\text{const.}}~\forall\, H_I $. It is because $\mathcal{D}^{\rightarrow}(\rho_{AB}|\rho_A^{\rm th})=0$ $\Rightarrow$ $[\rho_{AB}, \rho_A^{\rm th}]=0$ (see Ref.~\cite{discord}), which further implies that $[\rho_{AB},H_A ]=0$ when system A is a thermal state. The advertised result follows from Eq.~(\ref{gen-Crhiff}) and (\ref{frozen energy}). Similar results can be obtained by requiring system B to be thermal. These are summarized as follows.
\begin{align*}\label{frozen energy equilibrium}
&\mathcal{D}^{\rightarrow}(\rho_{AB}|\rho_A^{\rm th})=0 \Rightarrow  E_A={\rm const.}~\forall\, H_I \, , \\
&\mathcal{D}^{\leftarrow}(\rho_{AB}|\rho_B^{\rm th})=0   \Rightarrow  E_B={\rm const.}~\forall\, H_I \, , \\
&\mathcal{D}^{\leftrightarrow}(\rho_{AB}|\rho_A^{\rm th},\rho_B^{\rm th})=0  \Rightarrow E_A \, \& \, E_B={\rm const.} ~\forall\, H_I \, .
\end{align*}

\section{Bounding the power of energy flow}

Taking a step further, we shall show that the CANES also bounds the ``power" $\mathcal{P}_A \equiv|\frac{{d{E_A}}}{{dt}}|$ of energy flow between the system A and its environment B. From Eq.~(\ref{dEdttrtr}), we have explicitly $\mathcal{P}_A=\left|\tr \{ [H_I,H_A] (\rho_{AB}-\rho_{AB}^{A\textrm{in}}) \} \right|$  due to that $H_A$ and $\rho_{AB}^{A\rm{in}}$  commute, where
$\rho_{AB}^{A\rm{in}}\equiv(\Lambda_A\otimes \mathbb{I})\rho_{AB}$.
 By applying the H{\"o}lder's inequality, we obtain an upper bound (see Appendix \ref{upperbounds} for a complete derivation),
$\mathcal{P}_A \leqslant {\kappa_H} \left|\left| \rho_{AB}-\rho^{A\textrm{in}}_{AB}\right|\right|_1$, where ${\kappa_H} \equiv \left| \left| [H_I,H_A] \right| \right|_{\infty}$ is a state-independent constant that can be regarded as the coupling strength for energy flow, and the coherence-related part $\left|\left| \rho_{AB}-\rho^{A\textrm{in}}_{AB}\right|\right|_1$ can be bounded by the quantum Pinsker's inequality, which gives rise to
\begin{equation}\label{powerbound1}
\mathcal{P}_A \leqslant {\kappa_H}\sqrt {2\mathcal{C}^{\rightarrow}_e({\rho _{AB}})} \, .
\end{equation}
Note that $\mathcal{C}^{\rightarrow}_e ({\rho _{AB}}) = 0$ implies  $\mathcal{P}_A=0$, which is consistent with Theorem~\ref{thmfrezenlo}. In this sense, CANES may be regarded as a resource limiting the power of energy flow.

\section{CANES production in energy flow}

As CANES is indispensable for energy flow, we next investigate how much CANES is produced at each moment of time. For this purpose, we introduce the concept of \textit{effective} temperature for any non-equilibrium state as follows.
Given any quantum state $\rho$ and Hamiltonian $H$, the thermal state is given by $\rho^{\rm th}=e^{-\beta^{\rm th}H}/Z^{\rm th}$, where $Z^{\rm th}=\tr(e^{-\beta^{\rm th}H})$ is the partition function. The effective temperature, $1/\beta^{\rm th}$, of $\rho$ is defined by the condition where $\rho$ and $\rho^{\rm th}$ have the same average energy, i.e., $\tr(\rho H)=\tr(\rho^{\rm th} H)$.

Our goal is to investigate how much CANES is produced, i.e., the variation of CANES, $\Delta \mathcal{C}^{\leftrightarrow}_e$, when the effective temperatures of two systems A and B are different, i.e., $\beta _A^{\rm th} \ne \beta _B^{\rm th}$. Within an infinitesimal change of time, $\Delta t$, the whole system evolves from the state $\rho_{AB}$ to $\rho_{AB}'=\rho_{AB}+\Delta \rho_{AB}$.  For the case where the interaction energy $\left\langle {{H_I}} \right\rangle $ remains a constant in $\Delta t$ \cite{coupling energy}, the sum of the local energies are conserved, i.e., $\Delta E = \Delta E_B = - \Delta E_A$, which can be expressed (see Appendix \ref{expansions}) as,
\begin{equation}\label{DeltaE}
\Delta E = \frac{{{f_1}}}{{\beta _B^{\rm th} - \beta _A^{\rm th}}}\Delta t + \frac{{{f_2}}}{{\beta _B^{\rm th} - \beta _A^{\rm th}}}\Delta {t^2} + ...\, ,
\end{equation}
where $f_1 \equiv i\tr \{ H[\rho_{AB}, \ln (\rho^{\rm th}_A \otimes \rho^{\rm th}_B )]\}$ and $f_2 \equiv -\tr\{[H,\rho_{AB}][H,\ln (\rho^{\rm th}_A \otimes \rho^{\rm th}_B )]\}/2$. On the other hand, the short-time expansion of the CANES can be expressed as
\begin{equation}\label{DeltaC}
\Delta \mathcal{C}_e^ \leftrightarrow  = g_1 \Delta t + \left( {{g_2} - g_2^r} \right)\Delta {t^2} + ...\, ,
\end{equation}
where $g_1 \equiv i\tr\{H[\rho_{AB},\ln(\Lambda_A\otimes \Lambda_B)\rho_{AB}]\}$, $g_2 \equiv -\tr\{[H,\rho_{AB}][H,\ln (\Lambda_A\otimes \Lambda_B)\rho_{AB}]\}/2$, and $g^r_2 \equiv -\tr \{((\Lambda_A\otimes \Lambda_B)[H,\rho_{AB}])^2((\Lambda_A\otimes \Lambda_B)\rho_{AB})^{-1}\}/2$.

From Eq.~(\ref{DeltaE}) and (\ref{DeltaC}),  we can get  the variation relation between CANES production and energy flow:
\begin{equation}\label{varition-ralation}
(\beta^{\rm th}_B -\beta^{\rm th}_A )\Delta E=\eta \, \Delta \mathcal{C}^{\leftrightarrow}_e \, ,
\end{equation}
where $\eta=f_i/(g_j-g^r_j)$ with $f_i, g_j, g^r_j$ the minimal nonzero orders in the expansions (\ref{DeltaE}) and (\ref{DeltaC}),  which connects the energy flowed to the variations of CANES.
We perform the first and second order analysis for Eq.~(\ref{varition-ralation}) as follows.

(i) First-order analysis. Suppose the CANES and local energy (as functions of $t$) are not at stationary points simultaneously, i.e., $f_1 \neq 0$ or $g_1 \neq 0 $, we have 
\begin{equation}\label{f1/g1}
	\eta=f_1/g_1 \, ,
\end{equation}
and $\Delta E=d E/dt \Delta t, \Delta \mathcal{C}^\leftrightarrow_e=d \mathcal{C}^\leftrightarrow_e/dt \Delta t$ in Eq. (\ref{varition-ralation}).

Specially, when the systems A and B  are all in thermal equilibrium states, the entropy change $\Delta S_X = \beta^{\rm th}_{X}\Delta  E_X $. Thus the mutual information change $\Delta I_{AB}=\Delta S_A+ \Delta S_B-\Delta  S_{AB}=(\beta^{\rm th}_B-\beta^{\rm th}_A) \Delta  E$, where we have used energy conservation  $\Delta E=\Delta E_B=-\Delta E_A$, and  $\Delta S_{AB}=0$, by applying the identity $S\left( {U\rho {U^\dag }} \right) = S\left( \rho  \right)$. Together with (\ref{f1/g1}) one gets
  \begin{equation}\label{DeltaI}
  (\beta^{\rm th}_B-\beta^{\rm th}_A)\Delta E=\frac{f_1}{g_1} \Delta  \mathcal{C}^{\leftrightarrow}_e=\Delta I_{AB} \, .
  \end{equation}
It shows that for two thermal equilibrium systems with $f_1\neq 0$ or $g_1\neq 0$, the mutual information variation is just the energy flowed at any moment. In this process, the contribution of CANES is  $\Delta \mathcal{C}^{\leftrightarrow}/\Delta I_{AB}=g_1/f_1$.

(ii) Second-order analysis. When both the CANES and local energy are at stationary points, i.e., $f_1=g_1= 0$,
we have to consider the second order expansion. Typically, when the whole system in a zero CANES state,
i.e., $\mathcal{C}^{\leftrightarrow}_e(\rho_{AB})=0$, with $[H,\rho_{AB}]\neq 0$ and the local Hamiltonian $H_{A}$ and  $H_B$ being non-degenerate,  we have (see Appendix \ref{secondorder}) 
\begin{equation}\label{f2/g2}
	\eta=f_2/g_2 \, ,
\end{equation}
and $\Delta E=d^2 E/dt^2 \Delta t^2, \Delta \mathcal{C}^\leftrightarrow_e=d^2 \mathcal{C}^\leftrightarrow_e/dt^2 \Delta t^2$ in Eq. (\ref{varition-ralation}).


An important  case of (\ref{f2/g2}) is that the whole system is in a  product of two thermal equilibrium states, i.e.,  $\rho_{AB}=\rho^{\rm th}_A\otimes \rho^{\rm th}_B$. In this case, it is easy to check $f_2=g_2$ in Eq. (\ref{DeltaE}) and (\ref{DeltaC}), then $\eta =1$ from (\ref{f2/g2}).
Thus near a product of two thermal equilibrium states,  both the variation of CANES and the variation of energy  can be described by a kind of logarithmic skew information \cite{wigner} (see Appendix \ref{secondorder} and  \ref{logskewinformation}),
\begin{equation}\label{productstate}
(\beta^{\rm th}_B-\beta^{\rm th}_A)\Delta E=\Delta \mathcal{C}^{\leftrightarrow}_e=\mathcal{I}^s(H,\rho^{\rm th}_A\otimes \rho^{\rm th}_B)\Delta t^2 \, .
\end{equation}
The above relation shows that the variation of CANES is just the energy flowed, which emphasizes the significance of CANES in the very initial energy flow between two uncorrelated equilibrium states. 


\begin{figure}
\centering
\includegraphics[width=\columnwidth]{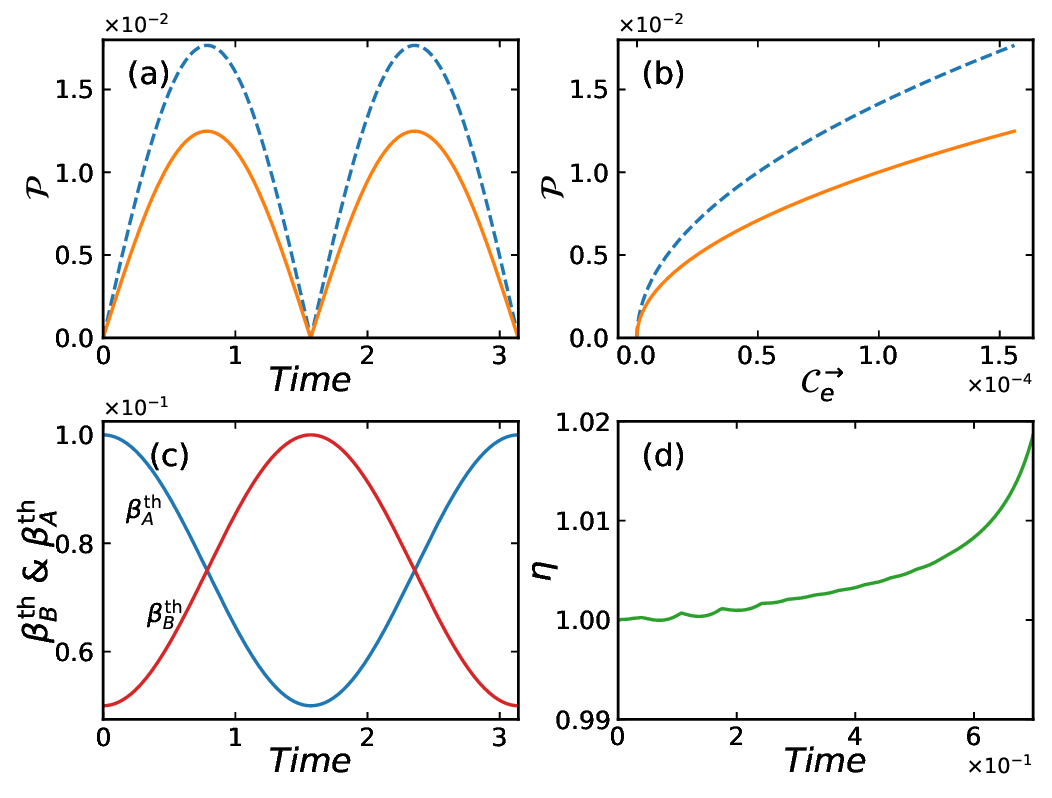}
\hspace{0.01in}
\caption{ Energy flow vs. CANES relation (color online).
  The subfigures (a) and (b) exhibit the power of energy flow (orange line) and its upper bound (\ref{powerbound1})  (blue dashed line) for the JC model between the atom and photon. The CANES limits the power of energy flow between the atom and photon.
  The subfigure (c) exhibits the inverse effective temperatures for the atom (blue line) and the light field (red line) respectively. The subfigure (d) exhibits the energy variation vs. CANES variation relations (\ref{productstate}) and (\ref{f1/g1}). At the very beginning of the evolution, through the second order analysis, the CANES variation is just the energy flowed, i.e., $\eta=f_2/g_2=1$.  For a later time, $\eta$ is determined by the first orders, i.e., $\eta=f_1/g_1$.
In this examples,  we set $\Delta=\omega/2$ (resonance case), $g=\omega$,  $\omega=1$, $\beta_A=0.1$ and $\beta_B=0.05$. These settings ensure that the total energy of the atom and light field is conserved.
}
\label{coherence_power_plot}
\end{figure}

\section{Example} 

A detailed example exhibiting the role played by CANES in energy flow is the Jaynes-Cummings (JC) model \cite{JCmodel} between a two-level atom and a single mode of photons. In the context of quantum thermodynamics, this model has been applied to quantum heat engine as well as light-harvesting complexes~\cite{Altintas}. The Hamiltonian of the JC model is given by $H_{JC}=H_A + H_B +H_I$. 
where $H_A=-\Delta \sigma_z$ and $H_B=\omega a^\dagger a$ are the local Hamiltonians for the  atom and light field respectively, $H_I=g(\sigma a^{\dagger}+\sigma^{\dagger}a)$ is the coupling between them, where $\sigma \equiv |g \ra \la  e|$. Initially, the system is assumed to be in a product of two thermal states with different temperatures,
\begin{equation}
	\rho_{AB}=e^{-\beta_A H_A}/Z_A\otimes e^{-\beta_B H_B}/Z_B \, ,
\end{equation}
where $Z_X=\tr (e^{-\beta_X H_X})$ is the partition function.

Fig. \ref{coherence_power_plot} shows the CANES as an upper bound for the power of energy flow, Eq. (\ref{powerbound1}), and  the CANES production versus energy variation, Eq. (\ref{f1/g1}) and (\ref{productstate}), in the energy exchange between the two level-atom and the light field.


\section{Conclusions}

We have introduced CANES to quantify the coherence among different energy eigen-subspaces of quantum states, and have presented its role played in energy flow. We proved that the local energy of any bipartite systems must be frozen if and only if the corresponding CANES vanishes. We have also demonstrated that the power of energy flow can be bounded by CANES. Furthermore, we have investigated the relations between the CANES production and energy flow. Our results show that the quantumness, in terms of the CANES, is an indispensable resource for energy flow.

Finally, we point out that for other physical quantities besides energy, e.g., $\la \mathcal{O} \ra$, which can be associated with  physical observable  $\mathcal{O}$, our approach is also suitable and similar results also hold. Coherence amnong $\mathcal{O}$'s eigen-subspaces is necessary in $\la \mathcal{O} \ra$'s flow, and the flow power is bounded by the coherence. This means that CANES like coherence with respect to a physical observable  is an indispensable resource in the flow of the physical quantity associate with that observable.

\bigskip
\noindent{\bf Acknowledgments}\, \, We thank Kavan Modi, Nai-Huan Jing, Lin Zhang, Yuan-Sheng Wang, and Zhi-Wen Liu for useful discussions. This work is supported by the NSFC under numbers
11401032 and
11675113, Beijing Municipal Commission of Education (KZ/201810028042),
China Scholarship Council (Grant No. 201808110022),
Natural Science Foundation of Guangdong Province
(2017B030308003), Guangdong Innovative and Entrepreneurial Research Team Program (2016ZT06D348),
and Science, Technology and Innovation Commission of Shenzhen
Municipality 
(ZDSYS201703031659262 and JCYJ20170412152620376).

\bigskip

\appendix

\section{\label{Proper_CANES}Properties of the CANES}

Under the definition of CANES, the set of incoherent states with respect to a Hamiltonian $H$ is given by
\begin{equation*}
	\mathcal{I}_e=\{ \rho | \rho=\Lambda \rho \},
\end{equation*}
where $\Lambda(\cdot)=\sum_i P_i (\cdot) P_i$ with $P_i$ being the projector onto the eigen-subspace of $H$ corresponding to the eigenvalue $h_i$. The incoherent operation $\mathcal{K}$ satisfies
\begin{equation*}
\mathcal{K}(\rho)\in \mathcal{I}_e, ~~\forall \rho \in \mathcal{I}_e.
\end{equation*}
With the above assumptions, following the the same steps as the ones given in  [Phys. Rev. Lett. \textbf{113}, 140401 (2014)], it is easy to show that the CANES (with relative entropy measure, i.e., the Eq. (1) in the main text) is a faithful measure of coherence, since (i) $\mathcal{C}_e(\rho)=0$ iff $\rho \in \mathcal{I}_e$; (ii) $\mathcal{C}_e(\rho)$ is convex, i.e., $\sum_i p_i \mathcal{C}_e(\rho_i) \geq \mathcal{C}_e(\sum_i p_i \rho_i)$, for $\sum_i p_i=1 ~\&~ p_i \geq 0$;
(iii) $\mathcal{C}_e(\rho)$ is non-increasing under incoherent operations, i.e., $\mathcal{C}_e(\rho)\geq \mathcal{C}_e(\mathcal{K}\rho)$.
Moreover, since CANES is within the definition of the relative entropy of superposition in [arXiv:quant-ph/0612146, (2006)], it also obeys the properties of relative entropy of superposition listed in [arXiv:quant-ph/0612146, (2006)].

\bigskip

Furthermore, we have the following properties for CANES. 

\smallskip
1. The CANES is actually the ``external'' coherence:
\begin{equation}\label{encop1}
\mathcal{C}_e(\rho)=C(\rho)-C(\rho^{\textrm{in}}),
\end{equation}
where $C(\rho)=S(\Pi \rho)-S(\rho)$ is the relative entropy coherence of state $\rho$ with respect to a set of basis,  $\Pi(\cdot)=\sum_i \Pi_i(\cdot)\Pi_i$ with the reference basis, $\{\Pi_i\}$, being a specific set of eigenbasis of $H$, the projected state $\rho^{\textrm{in}}\equiv \Lambda \rho=\sum_i P_i \rho P_i$ with $P_i$ being the projection onto the eigen-subspace of $H$ corresponding to the eigenvalue $h_i$.    $C(\rho^{\textrm{in}})$ only contains ``internal'' coherence, i.e.,  the coherence among different eigenbasis with the same eigenvalue of $H$.
\smallskip

\begin{proof}
Since $\{\Pi_i\}$ is a set of rank one projectors onto the  eigenbasis of $H$, we have $C(\rho)-C(\rho^{\textrm{in}})=S(\sum_i \Pi_i \rho \Pi_i)-S(\rho)-[S(\sum_i \Pi_i \rho^{\textrm{in}} \Pi_i)-S(\rho^{\textrm{in}})]=S(\rho^{\textrm{in}})-S(\rho)=S(\Lambda \rho)-S(\rho)=\mathcal{C}_e(\rho)$, where we have used $\sum_i \Pi_i \rho^{\textrm{in}} \Pi_i=\sum_i \Pi_i \rho \Pi_i$.
\end{proof}

\smallskip

Similar results like (\ref{encop1}) also hold for the bipartite CANES:

\begin{align}
	&\mathcal{C}^{\rightarrow}_e(\rho_{AB})=C^{\rightarrow}(\rho_{AB})-C^{\rightarrow}(\rho_{AB}^{A\textrm{in}}),\label{cel} \\
	&\mathcal{C}^{\leftarrow}_e(\rho_{AB})=C^{\leftarrow}(\rho_{AB})-C^{\leftarrow}(\rho_{AB}^{B\textrm{in}}),\label{cer}\\
	&\mathcal{C}^{\leftrightarrow}_e(\rho_{AB})=C^{\leftrightarrow}(\rho_{AB})-C^{\leftrightarrow}(\rho_{AB}^{AB\textrm{in}}),\label{cebi}
\end{align}
where $C^{\rightarrow}(\rho_{AB})=S(\Pi_A\otimes \mathbb{I}\rho_{AB})-S(\rho_{AB})$ with $\Pi_A(\cdot)=\sum_i \Pi^A_i (\cdot)\Pi^A_i$, is the left side relative entropy coherence of $\rho_{AB}$ with respect to a specific set of $H_A$'s eigenbasis $\{\Pi^A_i\}$,
$\rho_{AB}^{A\,\textrm{in}}\equiv (\Lambda_A \otimes \mathbb{I}) \rho_{AB}=\sum_i (P^A_i \otimes \mathbb{I})\rho_{AB} (P^A_i\otimes \mathbb{I})$, $\Lambda_A(\cdot)=\sum_i P^A_i(\cdot)P^A_i$ with $P^A_i$ being the projector onto the eigen-subspace of $H_A$ corresponding to the eigenvalue $h^A_i$ (the terms in Eq. (\ref{cer}) and (\ref{cebi}) are similar).

Note that although the values of total coherence $C(\rho)$ and internal coherence $C(\rho^{\textrm{in}})$ may have variation (when $H$ has degeneracy) due to  different chosen of the reference basis (eigenbasis of $H$), the CANES is independent of the basis chosen.  We  can also see that when $H$ is non-degenerate, $C(\rho^{\textrm{in}})=0$, then the CANES is reduced to the original relative entropy  coherence in [Phys. Rev. Lett. \textbf{113}, 140401 (2014)].

\bigskip
2. For the CANES, we have
\begin{equation}\label{encop2}
\mathcal{C}_{e}(\rho)=0 \Leftrightarrow \Lambda \rho=\sum_i P_i \rho P_i=\rho \Leftrightarrow  [\rho, H]=0,
\end{equation}
 where $P_i$ is the projector onto the eigen-subspace  of $H$ corresponding to the eigenvalue $h_i$.

\smallskip

\begin{proof}
$\mathcal{C}_{e}(\rho)=S(\Lambda\rho)-S(\rho)=0\Leftrightarrow \Lambda \rho=\sum_i P_i \rho P_i=\rho$. The Hamiltonian can be expressed as $H=\sum_i h_i P_i$. Then it is easy to check  $[H, \rho]=0$. Conversely, if $[H, \rho]=0$ with $H=\sum_i h_i P_i$, then
\begin{equation}\begin{aligned}
&[H, \rho]=0\\
=&[\sum_i h_i P_i, \sum_{ij} P_i \rho P_j]\\
=&\sum_{ij}h_i P_i \rho P_j- \sum_{ij} h_j P_i \rho P_j\\
=& \sum_{ij}(h_i-h_j)P_i \rho P_j.
\end{aligned}
\end{equation}
Since $h_i\neq h_j$ with $i\neq j$, we have $P_i \rho P_j=0$ for all $i\neq j$, which give rise to $\rho=\sum_{ij}P_i\rho P_j=\sum_i P_i \rho P_i=\Lambda \rho$.
\end{proof}

\smallskip

Similarly, for the left CANES, we have
\begin{equation}\begin{aligned}\label{leftCP2}
&\mathcal{C}^{\rightarrow}_{e}(\rho_{AB})=0\\
\Leftrightarrow& (\Lambda_A \otimes \mathbb{I}) \rho_{AB}=\sum_i (P^A_i \otimes \mathbb{I}) \rho_{AB} (P^A_i\otimes \mathbb{I})=\rho_{AB}\\
 \Leftrightarrow& [\rho_{AB}, H_A\otimes \mathbb{I}]=0.
\end{aligned}
\end{equation}
Note that when $H_A$ is non-degenerate, $\mathcal{C}^{\rightarrow}_{e}(\rho_{AB})=0$ implies that $\rho_{AB}$ can be written as an incoherent-quantum state, i.e.,
\begin{equation}
\rho_{AB}=\sum_i p_i |e^A_i\ra \la e^A_i|\otimes \rho^B_i,
\end{equation}
where $\{|e^A_i\ra\}$ is the eigenstates of $H_A$. This can be seen by
$\rho_{AB}=\sum_i (|e^A_i\ra \la e^A_i|\otimes \mathbb{I}) \rho_{AB}  (|e^A_i\ra \la e^A_i|\otimes \mathbb{I})$ from (\ref{leftCP2}).

\smallskip

For the right CANES,
\begin{equation}\begin{aligned}
&\mathcal{C}^{\leftarrow}_{e}(\rho_{AB})=0\\
 \Leftrightarrow& (\mathbb{I}\otimes \Lambda_B) \rho_{AB}= \sum_i (\mathbb{I}\otimes  P^B_i)  \rho_{AB} (\mathbb{I} \otimes P^B_i)=\rho_{AB}\\
\Leftrightarrow& [\rho_{AB},\mathbb{I}\otimes H_B]=0,
\end{aligned}
\end{equation}
and the bilateral CANES,
\begin{equation}\begin{aligned}\label{bilateralCP2}
&\mathcal{C}^{\leftrightarrow}_{e}(\rho_{AB})=0\\
\Leftrightarrow&(\Lambda_A\otimes \Lambda_B) \rho_{AB}=\sum_{ij} P^A_i \otimes P^B_j \rho_{AB} P^A_i\otimes P^B_j=\rho_{AB}\\
\Leftrightarrow&[\rho_{AB},\mathbb{I}\otimes H_B]=0 ~\&~[\rho_{AB},H_A\otimes \mathbb{I}]=0\\
\Leftrightarrow& \mathcal{C}^{\leftarrow}_{e}(\rho_{AB})=0~\&~\mathcal{C}^{\rightarrow}_{e}(\rho_{AB})=0.
\end{aligned}
\end{equation}
Eq. (\ref{bilateralCP2}) can be seen by
($\Lambda_A\otimes \Lambda_B) \rho_{AB}=\rho_{AB}\Leftrightarrow (\Lambda_A\otimes \mathbb{I}) \rho_{AB}=\rho_{AB}~\&~(\mathbb{I}\otimes \Lambda_B)\rho_{AB}=\rho_{AB}\Leftrightarrow [\rho_{AB},H_A\otimes \mathbb{I}]=0~\&~[\rho_{AB}, \mathbb{I}\otimes H_B]=0$. Note that when both $H_A$ and $H_B$ are non-degenerate, $\mathcal{C}^{\leftrightarrow}_{e}(\rho_{AB})=0$ implies that $\rho_{AB}$ can be written as an incoherent-incoherent (bipartite incoherent) state, i.e.,
\begin{equation}
\rho_{AB} =\sum_{ij} p_{ij} |e^A_i\ra \la e^A_i| \otimes  |e^B_j\ra \la  e^B_j|,
\end{equation}
where $|e^A_i\ra$s and $|e^B_j\ra$s are the  eigenbasis of $H_A$ and $H_B$, respectively. This can be seen by (\ref{bilateralCP2}).

\section{\label{changeingcorrelation}Changing quantum correlation with frozen local energies}

We give an example that the quantum correlation between two quantum systems A and B changes while local energies are frozen due to the degeneracy of local Hamiltonian.  The whole system evolves under Hamiltonian
\begin{equation*}
H=H_A\otimes \mathbb{I}+\mathbb{I}\otimes H_B+H_I,
\end{equation*}
where $H_A=10|0\ra \la 0|+ 10|1\ra \la 1|+2|2\ra \la 2|$, $H_B=5|0\ra \la 0|+5|1\ra \la 1|+2|2\ra \la 2|$, and the interaction $H_I=10|00\ra \la 10|+10|10\ra \la 00|$. Note that $H_X~(X=A,B)$ is degenerate in the  eigen-subspace $\rm{span}\{|0\ra, |1\ra \}$.

The whole system is initially in  state
\begin{equation*}
\rho_{AB}(0)=9/10|\psi\ra \la \psi| +1/10|22\ra \la 22|,
\end{equation*}
where $|\psi \ra=1/\sqrt{2}(|00\ra+|11\ra)$. The system  evolves  under the Liouville-von Neumann equation $\dot\rho_{AB}(t)=-i[H,\rho_{AB}(t)]$.

\begin{figure}
\centering
\includegraphics[width=2.8in]{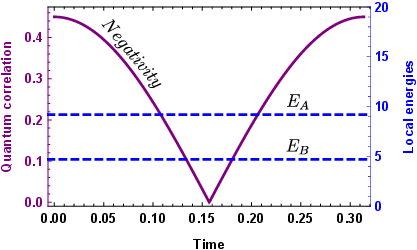}
\caption{
The quantum correlation (purple line) between two quantum systems A and B changes while the local energies (blue dashed lines) are frozen. We use the entanglement measurement negativity, $\mathcal{N}(\rho_{AB})=1/2(||\rho_{AB}^{T_A}||_1-1)$, to capture the quantum correlation between the systems A and B, where $\rho^{T_A}$ is the partial transpose of $\rho$, and $||\cdot||_1$ is the trace norm.}\label{fig1}
\end{figure}

It is easy to check that $[\rho_{AB}(t),H_X\otimes \mathbb{I}]=0~\forall t\geq 0$, i.e., the corresponding CANES is zero. Then our theorem 1 assures that the energies of the local systems keep a constant during the evolution, i.e., there is no energy flow between systems A and B. However, the entanglement (including quantum discord) between systems A and B changes over the evolution (see Fig. \ref{fig1}).

\section{\label{upperbounds}The upper bounds of the power of energy flow}

From Theorem 1 in the main text, by straightforward calculations we get the first derivative of energy,
\begin{equation}\begin{aligned}\label{bound0}
\frac{dE_A}{dt}&=-i \tr \{  H_I[\rho_{AB},H_A] \} \\
&=-i \tr \{  H_I[\rho_{AB}-\rho_{AB}^{A\rm{in}},H_A] \} \\
&=i \tr \{ [H_I,H_A] (\rho_{AB}-\rho_{AB}^{A\textrm{in}}) \},\\
\end{aligned}
\end{equation}
due to that $H_A$ and $\rho_{AB}^{A\rm{in}}$  commute, where
$\rho_{AB}^{A\rm{in}}\equiv (\Lambda_A\otimes \mathbb{I}) \rho_{AB}=\sum_i(P^A_i\otimes \mathbb{I})\rho_{AB} (P^A_i\otimes \mathbb{I})$ with $P^A_i$ the projector onto the eigen-subspace of $H_A$ corresponding to the eigenvalue $h^A_i$.

Then the power of energy flow, $\mathcal{P}_A \equiv \left|\frac{d E_A}{dt} \right|$, satisfies
\begin{equation}\label{power1}
\begin{aligned}
\mathcal{P}_A&=\left|\tr \{ [H_I, H_A] (\rho_{AB}-\rho_{AB}^{A\textrm{in}}) \}\right|\\
&=\left|\sum_{i} {[H_I, H_A]}_{ii} {(\rho_{AB}-\rho_{AB}^{A\textrm{in}})}_i \right| \\
&\leq \sum_i \left|{[H_I, H_A]}_{ii} {(\rho_{AB}-\rho_{AB}^{A\textrm{in}})}_i \right| \\
&\leq \left|\left| [H_I,H_A]_\textrm{d}  \right|\right|_{\infty} \left|\left| \rho_{AB}-\rho_{AB}^{A\textrm{in}}\right|\right|_1,
\end{aligned}
\end{equation}
where $[H_I,H_A]_{ii}=\la \psi_i|[H_I,H_A]|\psi_i\ra$ are the diagonal elements of $[H_I,H_A]$ under the eigenstates of $(\rho_{AB}-\rho_{AB}^{A\rm{in}})=\sum_i (\rho_{AB}-\rho_{AB}^{A\rm{in}})_i |\psi_i\ra \la \psi_i|$, with $(\rho_{AB}-\rho_{AB}^{A\rm{in}})_i$ and $|\psi_i\ra$ being the eigenvalues and eigenstates respectively, $[H_I,H_A]_{\rm d}=\sum_i [H_I,H_A]_{ii} |\psi_i\ra \la \psi_i|$ (here the subscript `$\rm d$' denotes the projection with respect to basis $\{|\psi_i\ra\}$).
The last inequality is due to the H{\"o}lder's inequality: for two vectors $X$ and $Y$, $|\la X| Y \ra|\leq ||X||_\infty ||Y||_1$ with $||\cdot||_\infty$ and $||\cdot||_1$ being the infinity norm and trace norm respectively.

Owning to $||A_{\rm d}||_\infty \leq ||A||_\infty$ for a matrix $A$ (a projective measurement decreases the infinity norm), and the quantum Pinsker's inequality $S(\rho||\sigma)\geq \frac{1}{2}\left|\left|\rho-\sigma\right|\right|^2_1$ for two states $\rho$ and $\sigma$, we have 
\begin{equation}
	\begin{aligned}\label{power2}
	\mathcal{P}_A&\leq \left|\left| [H_I, H_A]_\textrm{d} \right|\right|_{\infty} \sqrt{2 S(\rho_{AB}||\rho_{AB}^{A\textrm{in}})}\\
&= \left|\left| [H_I, H_A]_\textrm{d} \right|\right|_{\infty} \sqrt{2 \mathcal{C}_e^{\rightarrow}(\rho_{AB})}  \\
&\leq  \left|\left| [H_I, H_A] \right|\right|_{\infty} \sqrt{2 \mathcal{C}_e^{\rightarrow}(\rho_{AB})}
	\end{aligned}
\end{equation}
from (\ref{power1}), where we have used $S(\rho_{AB}||\rho_{AB}^{A\textrm{in}})=S(\rho_{AB}^{A\textrm{in}})-S(\rho_{AB})=\mathcal{C}^{\rightarrow}_e(\rho_{AB})$. There is an equality  in (\ref{power2}) when $\mathcal{C}_e^{\rightarrow}(\rho_{AB})$ gets zero.

\section{\label{expansions}Expansions for energy flow and CANES}

For the energy flow, we have
\begin{equation}\begin{aligned}\label{DE}
&(\beta^{\rm th}_B -\beta^{\rm th}_A )\Delta E\\
=&-\tr(\Delta \rho_A \ln \rho^{\rm th}_A )-\tr(\Delta \rho_B \ln \rho^{\rm th}_{B})\\
=&-\tr [ \Delta \rho_{AB}\ln (\rho^{\rm th}_A \otimes \rho^{\rm th}_{B})],
\end{aligned}
\end{equation}
where we have used  $\Delta E_{X}=\tr(\Delta \rho_{X} H_{X})$~$(X=A,B)$,  $\Delta E=\Delta E_B=-\Delta E_A$ (energy conservation),  $\ln {\rho^{\rm th}_{X}}=-\beta^{\rm th}_{X} H_{X}-\ln Z^{\rm th}_{X}$, and  $\Delta \rho_{A(B)}=\tr_{B(A)}(\Delta \rho_{AB})$.
The evolution of the whole system is governed by the Liouville-von Neumann equation,
$\dot\rho_{AB}=-i[H,\rho_{AB}]$,
where $i=\sqrt{-1}$ and $[a,b]=ab-ba$. By Taylor's formula, we can expand $\rho_{AB}$ in terms of $\Delta t$,
\begin{equation}\label{Deltarhoab}
\Delta \rho_{AB}=-i[H,\rho_{AB}]\Delta t-\frac{1}{2}[H,[H,\rho_{AB}]]\Delta t^2+....
\end{equation}
Plugging (\ref{Deltarhoab}) into  (\ref{DE}) and  using the cycle property of the trace, we have the expansion for the energy flow in terms of $\Delta t$,
\begin{equation}\label{DES}
(\beta^{\rm th}_B -\beta^{\rm th}_A )\Delta E=f_1\Delta t+f_2 \Delta t^2+...,
\end{equation}
where
\begin{equation}\label{fs}
\begin{aligned}
&f_1=i\tr \{ H[\rho_{AB}, \ln (\rho^{\rm th}_A \otimes \rho^{\rm th}_B )]\},\\
&f_2=-\tr\{[H,\rho_{AB}][H,\ln (\rho^{\rm th}_A \otimes \rho^{\rm th}_B )]\}/2, \\
\end{aligned}
\end{equation}
which  correspond to the first derivative of energy, $d E/dt=f_1/(\beta^{\rm th}_B -\beta^{\rm th}_A )$, and  the second derivative of energy, $d^2 E/dt^2=2 f_2/(\beta^{\rm th}_B -\beta^{\rm th}_A )$, respectively.

For the variation of CANES, we have
\begin{equation}\begin{aligned}\label{DC}
\Delta \mathcal{C}^{\leftrightarrow}_e=\Delta (S_{\widetilde{A}\widetilde{B}}-S_{AB})=\Delta S_{\widetilde{A}\widetilde{B}},
\end{aligned}
\end{equation}
here  $\widetilde{X}$ $(X=A, B)$  denotes the projection  on the system $X$, e.g., $\rho_{\widetilde{A}\widetilde{B}}=(\Lambda_A\otimes \Lambda_B) \rho_{AB}$, and  $S_{\widetilde{A}\widetilde{B}}=S(\rho_{\widetilde{A}\widetilde{B}})$. In the last equality we have used $\Delta S_{AB}=0$  due to that the evolution of the whole system is unitary. By straightforward calculations, we have
\begin{equation}\label{DEC}
\begin{aligned}
&\Delta S(\rho_{\widetilde{A}\widetilde{B}})\\
=&-\tr(\Delta \rho_{\widetilde{A}\widetilde{B}} \ln \rho_{\widetilde{A}\widetilde{B}})-S(\rho_{\widetilde{A}\widetilde{B}}'||\rho_{\widetilde{A}\widetilde{B}})\\
=&-\tr \{\Delta \rho_{AB}\ln (\Lambda_A\otimes \Lambda_B)\rho_{AB}\}-S(\rho_{\widetilde{A}\widetilde{B}}'||\rho_{\widetilde{A}\widetilde{B}}),
\end{aligned}
\end{equation}
where we have used the fact that $\Delta$ and $\Lambda_X$ are commutating (since $\Lambda_X$ is fixed), $\Lambda \ln \Lambda \rho=\ln \Lambda\rho$ for any projection $\Lambda$, and $\rho_{\widetilde{A}\widetilde{B}}'=(\Lambda_A\otimes \Lambda_B) \rho_{AB}'$.
For the relative entropy term in (\ref{DEC}), one can find that the first order of $\Delta t$ in its Taylor's expansion vanishes ($g^r_1=0$), and
\begin{equation}\label{DECR}
S(\rho_{\widetilde{A}\widetilde{B}}'||\rho_{\widetilde{A}\widetilde{B}})=g^r_2 \Delta t^2+...,
\end{equation}
where $g^r_2=\tr \{(\dot{\rho}_{\widetilde{A}\widetilde{B}})^2 (\rho_{\widetilde{A}\widetilde{B}})^{-1} \} /2=-\tr \{ ((\Lambda_A\otimes \Lambda_B)[H,\rho_{AB}])^2((\Lambda_A\otimes \Lambda_B)\rho_{AB})^{-1} \} /2$.
Substituting (\ref{Deltarhoab}) and (\ref{DECR}) into (\ref{DEC}), and taking into account the relation (\ref{DC}),
we have the expansion for the variation of CANES  in terms of $\Delta t$,
\begin{equation*}\label{DECS}
\Delta \mathcal{C}^{\leftrightarrow}_e= g_1\Delta t+(g_2-g^r_2) \Delta t^2+...,
\end{equation*}
where
\begin{equation*}\label{gs}
\begin{aligned}
&g_1=i\tr\{H[\rho_{AB},\ln(\Lambda_A\otimes \Lambda_B)\rho_{AB}]\},\\
&g_2=-\tr\{[H,\rho_{AB}][H,\ln (\Lambda_A\otimes \Lambda_B)\rho_{AB}]\}/2,\\
&g^r_2=-\tr \{((\Lambda_A\otimes \Lambda_B)[H,\rho_{AB}])^2/((\Lambda_A\otimes \Lambda_B)\rho_{AB})\}/2,
\end{aligned}
\end{equation*}
which  correspond to the first  derivative of the CANES, $d \mathcal{C}^{\leftrightarrow}_e/dt=g_1$, and the second derivative of the CANES, $d^2 \mathcal{C}^{\leftrightarrow}_e/dt^2=2(g_2-g^r_2)$, respectively.

\section{\label{secondorder}The second order analysis of CANES production versus energy flow}

For a zero CANES state $\rho_{AB}$, we have $\mathcal{C}^{\leftrightarrow}_e(\rho_{AB})=0 \Leftrightarrow (\Lambda_A\otimes \Lambda_B)\rho_{AB}=\rho_{AB}$, which leads to  $g_1=i\tr\{H[\rho_{AB},\ln(\Lambda_A\otimes \Lambda_B)\rho_{AB}]\}=0$.
As the product state
\begin{equation*}\begin{aligned}
&\rho_A^{\rm th}\otimes \rho_B^{\rm th}\\
\propto& e^{-\beta^{\rm th}_A H_A} \otimes  e^{-\beta^{\rm th}_B H_B}\\
=&e^{-(\beta^{\rm th}_A H_A\otimes \mathbb{I}+\beta^{\rm th}_B\mathbb{I}\otimes H_B)},
\end{aligned}
 \end{equation*}
we have $[\rho_{AB},\ln (\rho_A^{\rm th}\otimes \rho_B^{\rm th})]=-[\rho_{AB},\beta^{\rm th}_A H_A\otimes \mathbb{I}+\beta^{\rm th}_B\mathbb{I}\otimes H_B]$. On the other hand, we have $\mathcal{C}^{\leftrightarrow}_e(\rho_{AB})=0 \Leftrightarrow  [\rho_{AB},\mathbb{I}\otimes H_B]=0 ~ \& ~ [\rho_{AB},H_A\otimes \mathbb{I}]=0 \Rightarrow  [\rho_{AB},\beta^{\rm th}_A H_A\otimes \mathbb{I}+\beta^{\rm th}_B\mathbb{I}\otimes H_B]=0$, which leads to $f_1=i\tr \{ H[\rho_{AB}, \ln (\rho^{\rm th}_A \otimes \rho^{\rm th}_B )]\}=0$. Hence we need to consider the second order.

For the term $g^r_2$, assuming that $H_{A}$ and $H_B$ are non-degenerate,  we have $\rank (P^{A(B)}_i)=1 \, \forall i$. It is easy to check $(\Lambda_A\otimes \Lambda_B)[H,\rho_{AB}]=0$,
which leads to $g^r_2=-\tr \{((\Lambda_A\otimes \Lambda_B)[H,\rho_{AB}])^2/((\Lambda_A\otimes \Lambda_B)\rho_{AB})\}/2=0$.
For the term $g_2$, from $(\Lambda_A\otimes \Lambda_B)\rho_{AB}=\rho_{AB}$, we obtain
\begin{equation*}\begin{aligned}
g_2=&-\tr\{[H,\rho_{AB}][H,\ln (\Lambda_A\otimes \Lambda_B)\rho_{AB}]\}/2\\
=&-\tr\{[H,\rho_{AB}][H,\ln \rho_{AB}]\}/2.
\end{aligned}
\end{equation*}
Given  $[H,\rho_{AB}]\neq0$, one can show that the above $g_2=\mathcal{I}^s(H,\rho_{AB})$, which  is always positive (see Appendix \ref{logskewinformation}).
Finally, we have $f_1=g_1=g^r_1=g^r_2=0$, and  $g_2 \neq 0$, which means that the second order expansion is enough to determine  $\eta=f_2/g_2$.

Specifically, when the state is a product of two thermal state, i.e., $\rho_{AB}=\rho^{\rm{th}}_A\otimes \rho^{\rm{th}}_B$, we have $\eta=1$ due to $f_2=g_2=\mathcal{I}^s(H,\rho^{\rm{th}}_A \otimes \rho^{\rm{th}}_B)$. In this case, we have 
$(\beta^{\rm th}_B-\beta^{\rm th}_A)\Delta E=\Delta \mathcal{C}^{\leftrightarrow}_e=\mathcal{I}^s(H,\rho^{\rm{th}}_A \otimes \rho^{\rm{th}}_B)$.

\section{\label{logskewinformation}The logarithmic skew information}

The logarithmic skew information $\mathcal{I}^s(H,\rho)=-1/2\tr \{[\rho,H][\ln \rho, H]\}\geq 0$, with the equality holding if and only if $[\rho, H]=0$.

\begin{proof}
Under $\rho$'s eigenbasis $|i\ra$, we can write $\rho=\sum_i p_i |i\ra \la i|$ and $H=\sum_{ij}h_{ij}|i\ra \la j|$. Hence
$[\rho,H]=\sum_{ij} h_{ij} (p_i-p_j)|i\ra \la j|,$
and
$[\ln \rho,H]=\sum_{kl} h_{kl} (\ln p_k-\ln p_l)|k\ra \la l|.$
Thus
\begin{equation*}\begin{aligned}
\mathcal{I}^s(H,\rho)&=-\frac{1}{2}\tr \{\sum_{ijl}  h_{ij} (p_i-p_j)  h_{jl} (\ln p_j-\ln p_l) |i\ra \la l|\}\\
&=\frac{1}{2} \sum_{ij}  |h_{ij}|^2 (p_i-p_j) (\ln p_i-\ln p_j).
\end{aligned}\end{equation*}
Due to monotone property of  log function, $(p_i-p_j) (\ln p_i-\ln p_j)\geq 0~\forall i,j$, we have $\mathcal{I}^s(\rho,H)\geq 0$, with the equality holding if and only if $|h_{ij}|^2 (p_i-p_j) (\ln p_i-\ln p_j)=0, \,\forall i,j$. This requires that if $p_i\neq p_j$, then $h_{ij}=0$, which means that $H$ is diagonal under $\rho$'s all non-degenerate eigen-subspaces, i.e., $[\rho, H]=0$.
\end{proof}


\begin{thebibliography}{unsrt}
\bibitem{coupling energy} To make sure the change of interaction energy being zero for any state, one can adpot the condition  $[H, H_A\otimes I + I \otimes  H_B]=0$. However, given the variation of interaction energy is relatively small in $\Delta t$, our results also holds without this condition.
\end{thebibliography}
\end{document}